\documentclass[12pt]{article}
\usepackage{graphicx}

\begin{document}

\title{\textbf{Fusion dynamics of symmetric systems near barrier energies}}

\author{Zhao-Qing Feng$^{a,b}$\footnote{Corresponding author. Tel. +86 931 4969215. \newline \emph{E-mail address:} fengzhq@impcas.ac.cn},
Gen-Ming Jin$^{a,b}$}
\date{}
\maketitle

\begin{center}
$^{a}${\small \emph{Institute of Modern Physics, Chinese Academy of
Sciences, Lanzhou 730000, China}}
\end{center}

\textbf{Abstract}
\par
The enhancement of the sub-barrier fusion cross sections was
explained as the lowering of the dynamical fusion barriers within
the framework of the improved isospin-dependent quantum molecular
dynamics (ImIQMD) model. The numbers of nucleon transfer in the neck
region are appreciably dependent on the incident energies, but
strongly on the reaction systems. A comparison of the neck dynamics
is performed for the symmetric reactions $^{58}$Ni+$^{58}$Ni and
$^{64}$Ni+$^{64}$Ni at energies in the vicinity of the Coulomb
barrier. An increase of the ratios of neutron to proton in the neck
region at initial collision stage is observed and obvious for
neutron-rich systems, which can reduce the interaction potential of
two colliding nuclei. The distribution of the dynamical fusion
barriers and the fusion excitation functions are calculated and
compared them with the available experimental data.

\bigskip

Heavy-ion fusion reactions at energies in the vicinity of the
Coulomb barrier has been an important subject in nuclear physics for
more than 20 years, which is involved in not only exploring several
fundamental problems such as quantum tunneling in the
multidimensional potential barrier etc, also investigating nuclear
physics itself associated with nuclear structure, synthesis of
superheavy nuclei etc $\cite{Ba98}$. The experimental fusion cross
sections can be well reproduced by the various coupled channel
methods, which include the couplings of the relative motion to the
nuclear shape deformations, vibrations, rotations, and
nucleon-transfer, such as CCFULL code $\cite{Ha99}$. However, the
coupled channel models still have some difficulties in describing
the fusion reactions for symmetric systems, especially for heavy
combinations, in which the neck dynamics in the fusion process of
two colliding nuclei plays an important role on the interaction
potential, and consequently on the fusion cross section. Microscopic
mechanism of the neck dynamics is significant for properly
understanding the capture and fusion process in the formation of
superheavy nuclei in massive fusion reactions $\cite{Fe07}$. The
ImIQMD model has been successfully applied to treat heavy-ion fusion
reactions near barrier energies in our previous works $\cite{Fe05}$,
in which the interaction potential energy is microscopically derived
from the Skyrme energy-density functional besides the spin-orbit
term and the shell correction is considered properly. In this
letter, we will concentrate on exploring the influence of the
dynamical mechanism in heavy-ion collisions near barrier energies on
the fusion cross sections.

In the ImIQMD model, the time evolutions of the nucleons under the
self-consistently generated mean-field are governed by Hamiltonian
equations of motion, which are derived from the time dependent
variational principle and read as
\begin{eqnarray}
\dot{\mathbf{p}}_{i}=-\frac{\partial H}{\partial\mathbf{r}_{i}},
\quad \dot{\mathbf{r}}_{i}=\frac{\partial
H}{\partial\mathbf{p}_{i}}.
\end{eqnarray}
The total Hamiltonian $H$ consists of the kinetic energy, the
effective interaction potential and the shell correction part as
\begin{equation}
H=T+U_{int}+U_{sh}.
\end{equation}
The details of the three terms can be found in details in Ref.
$\cite{Fe05}$. The shell correction term is important for magic
nuclei induced fusion reactions, which constrains the fusion cross
section in the sub-barrier region.

For the lighter reaction systems, the compound nucleus is formed
after the two colliding nuclei is captured by the interaction
potential. The quasi-fission reactions after passing over the
barrier take place when the product $Z_{p}Z_{t}$ of the charges of
the projectile and target nuclei is larger than about 1600. In the
ImIQMD model, the interaction potential $V(R)$ of two colliding
nuclei as a function of the distance $R$ between their centers is
defined as $\cite{Br68}$
\begin{equation}
V(R)=E_{pt}(R)-E_{p}-E_{t}.
\end{equation}
Here the $E_{pt}$, $E_{p}$ and $E_{t}$ are the total energies of the
whole system, projectile and target, respectively. The total energy
is the sum of the kinetic energy, the effective potential energy and
the shell correction energy. In the calculation, the Thomas-Fermi
approximation is adopted for evaluating the kinetic energy. Shown in
in Fig. 1 is a comparison of the various static interaction
potentials, such as Bass potential $\cite{Ba77}$, double-folding
potential used in dinuclear system model $\cite{Fe07}$, proximity
potential of Myers and Swiatecki $\cite{My00}$, the adiabatic
barrier as mentioned in Ref. $\cite{Si01}$ and ImIQMD static and
dynamical interaction potentials for head on collisions of the
reaction system $^{58}$Ni+$^{58}$Ni. It should be noted that the
potentials calculated by the ImIQMD model have included the shell
effects that evolve from the projectile and target nuclei into the
composite system. The contribution of the shell correction energy to
the interaction potential is shown separately in the right panel of
the figure at frozen densities and different incident energies. The
static interaction potential means that the density distribution of
projectile and target is always assumed to be the same as that at
initial time, which is a diabatic process and depends on the
collision orientations and the mass asymmetry of the reaction
systems. The corresponding barrier heights are indicated for the
various cases. However, for a realistic heavy-ion collision, the
density distribution of the whole system will evolve with the
reaction time, which is dependent on the incident energy and impact
parameter of the reaction system $\cite{Fe05a}$. In the calculation
of the dynamical potentials, we only pay attention to the fusion
events, which give the dynamical fusion barrier. At the same time, a
stochastic rotation is performed for each simulation event. One can
see that the heights of the dynamical barriers are reduced gradually
with decreasing the incident energy, which result from the
reorganization of the density distribution of two colliding nuclei
due to the influence of the effective interaction potential on each
nucleon. The dynamical barrier with incident energy $E_{c.m.}$=105
MeV approaches the static one. The lowering of the dynamical fusion
barrier is in favor of the enhancement of the sub-barrier fusion
cross sections, which can give a little information that the cold
fusion reactions are also suitable to produce superheavy nuclei
although an extra-push energy is needed for heavy reaction systems
$\cite{Sw82}$. The energy dependence of the nucleus-nucleus
interaction potential in heavy-ion fusion reactions was also
investigated by the time dependent Hartree-Fock theory and the
lowering of dynamical barrier near Coulomb energies was also
observed $\cite{Wa08}$.

%%%%%%%%%%%%%%%%%%%%%%%%%%%% Fig.1 here %%%%%%%%%%%%%%%%%%%%%%%%%%%%%%%%%

The influence of the structure quantities such as excitation
energies, deformation parameters of the collective motion can be
embodied by comparing the fusion barrier distributions calculated
from the coupled channel models and the measured fusion excitation
functions. In the ImIQMD model, the dynamical fusion barrier is
calculated by averaging the fusion events at a given incident energy
and a fixed impact parameter. To explore more information on the
fusion dynamics, we also investigate the distribution of the
dynamical fusion barrier, which counts the dynamical barrier per
fusion event and satisfies the condition $\int
f(B_{fus})dB_{fus}=1$. Fig. 2 shows the barrier distribution for
head on collisions of the reaction $^{58}$Ni+$^{58}$Ni at the center
of mass incident energies 96 MeV and 100 MeV, respectively, which
correspond to below and above the static barrier $V_{b}=97.32$ MeV
as labeled in Fig. 1, and a comparison with the neutron-rich system
$^{64}$Ni+$^{64}$Ni. The distribution trend moves towards the
low-barrier region with decreasing the incident energy, which can be
explained from the slow evolution of the colliding system. The
system has enough time to exchange and reorganize nucleons of the
reaction partners at lower incident energies. A number of fusion
events are located at the sub-barrier region, which is favorable to
enhance sub-barrier fusion cross sections. There is a little
distribution probability that the fusion barrier is higher than the
incident energy 96 MeV owing to dynamical evolution of two touching
nuclei. We should note that the fusion events decrease dramatically
with incident energy in the sub-barrier region. Neutron-rich system
has the distribution towards the low-barrier region owing to the
lower dynamical fusion barrier, which favors the enhancement of the
fusion cross section.

%%%%%%%%%%%%%%%%%%%%%%%%%%%% Fig.2 here %%%%%%%%%%%%%%%%%%%%%%%%%%%%%%%%%

The neck formation in heavy-ion collisions close to the Coulomb
barrier is of importance for understanding the enhancement of the
sub-barrier cross sections. A phenomenological approach (neck
formation fusion model) was proposed by Vorkapi\'{c} $\cite{Vo94}$
to fit experimental data that can not be reproduced properly by the
coupled channel models. Using a classical dynamical model Aguiar,
Canto, and Donangelo have pointed out that the neck formation in
heavy-ion fusion reactions may explain the lowering of the barrier
$\cite{Ag85}$. Using the ImIQMD model, we carefully investigate the
dynamics of the formation of the neck in heavy-ion fusion reactions.
The neck region is defined as a cylindrical shape along the
collision orientation with the high 4 fm when the density at the
touching point reaches 0.02$\rho_{0}$. Shown in Fig. 3 is the
numbers of nucleon transfer from projectile to target in the neck
region at incident energies 95 MeV and 100 MeV in the left panel and
a comparison of the system $^{58}$Ni+$^{58}$Ni and
$^{64}$Ni+$^{64}$Ni in the right panel. The evolution time starts at
the stage of the neck formation. A slight peak appears for both
cases because the dynamical fluctuation takes place in the formation
process of the neck. Larger numbers of neutron transfer are obvious
especially for neutron-rich system, which can be easily understood
because the neutron transfer does not affected by the repulsive
Coulomb force. The transfer of protons reduces the interaction
potential of two colliding nuclei. The time evolution of the ratio
of neutron to proton in the neck region and the radius of the neck
at incident energy 100 MeV are also calculated as shown in Fig. 4
for the reactions $^{58}$Ni+$^{58}$Ni and $^{64}$Ni+$^{64}$Ni. It is
clear that the neutron-rich system has the larger values of the N/Z
ratio and the neck radius. An obvious bump in the evolution of the
N/Z ratio appears at the initial stage of the formation of the neck
for both systems due to the Coulomb repulsion for protons.

%%%%%%%%%%%%%%%%%%%%%%%%%%%% Fig.3 here %%%%%%%%%%%%%%%%%%%%%%%%%%%%%%%%%
%%%%%%%%%%%%%%%%%%%%%%%%%%%% Fig.4 here %%%%%%%%%%%%%%%%%%%%%%%%%%%%%%%%%

In the ImIQMD model, the fusion cross section is calculated by the
formula $\cite{Fe05}$
\begin{equation}
\sigma_{fus}(E)=2\pi\int_{0}^{b_{max}}bp_{fus}(E,b)db=2\pi\sum_{b=\Delta
b}^{b_{max}}bp_{fus}(E,b)\Delta b,
\end{equation}
where $p_{fus}(E,b)$ stands for the fusion probability and is given
by the ratio of the fusion events $N_{fus}$ to the total events
$N_{tot}$. In the calculation, the step of the impact parameter is
set to be $\Delta b=0.5$ fm. In Fig. 5 we show a comparison of the
calculated fusion excitation functions and the well-known one
dimensional Hill-Wheeler formula $\cite{Hi53}$ as well as the
experimental data for the reactions $^{58}$Ni+$^{58}$Ni
$\cite{Be81}$ and $^{64}$Ni+$^{64}$Ni $\cite{Ji04}$. One can see
that a strong enhancement of the fusion cross sections for the
neutron-rich combination $^{64}$Ni+$^{64}$Ni is obvious, especially
in the sub-barrier region. The Hill-Wheeler formula reproduces
rather well the fusion cross sections at above barrier energies, but
underestimate obviously the sub-barrier cross sections. The ImIQMD
model reproduces the experimental data rather well over the whole
range. In the piont of view from dynamical calculations, the
reorganization of the density distribution of the colliding system
results in the lowering of the dynamical fusion barrier, which
consequently enhances the sub-barrier fusion cross sections. The
phenomenon is more clearly for neutron-rich combinations.

%%%%%%%%%%%%%%%%%%%%%%%%%%%% Fig.5 here %%%%%%%%%%%%%%%%%%%%%%%%%%%%%%%%%

In conclusion, using the ImIQMD model, the fusion dynamics in
heavy-ion collisions in the vicinity of the Coulomb barrier is
investigated systematically. The dynamical fusion barrier is reduced
with decreasing the incident energies, which results in the
enhancement of the sub-barrier fusion cross sections. The
distribution forms of the dynamical fusion barrier are dependent on
the incident energies and the N/Z ratios in the neck region of the
reaction systems. The nucleon transfer in the neck region reduces
the interaction potential of two colliding nuclei. The lower fusion
barrier is in favor of the enhancement of the fusion cross sections
of the neutron-rich systems.

\textbf{Acknowledgements}

This work was supported by the National Natural Science Foundation
of China under Grant No. 10805061, the special foundation of the
president fellowship, the west doctoral project of Chinese Academy
of Sciences, and major state basic research development program
under Grant No. 2007CB815000.

\newpage
%%%%%%%%%%%%%%%%%%%%%%%%%%%%%%%%%%% figure 1 %%%%%%%%%%%%%%%%%%%%%%%%%%%%
\begin{figure}
\begin{center}
{\includegraphics*[width=0.8\textwidth]{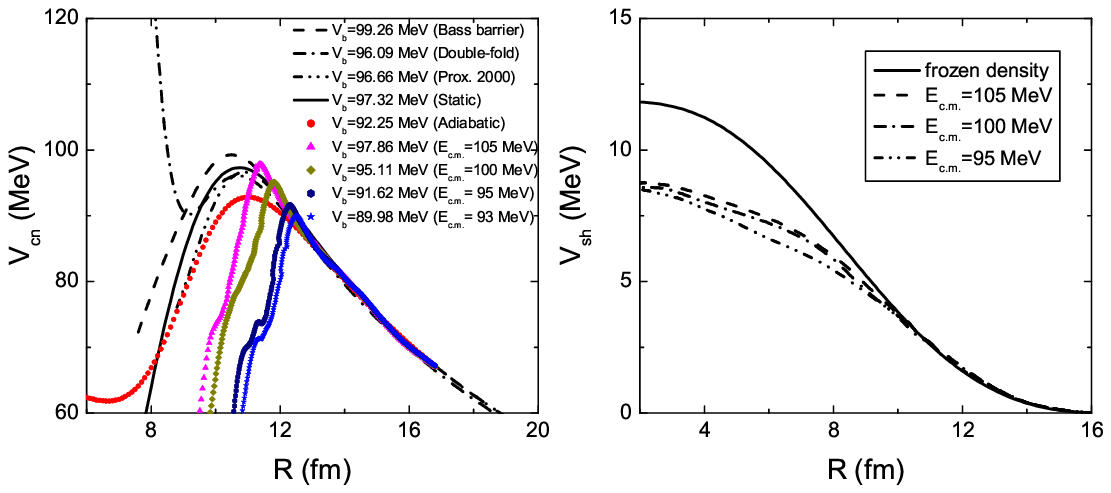}}
\end{center}
\caption{Comparisons of the reaction $^{58}$Ni+$^{58}$Ni for various
static interaction potentials (Bass, double-folding, proximity and
ImIQMD potential at frozen density), the dynamical fusion potentials
at different incident energies and the adiabatic potential in Ref.
$\cite{Si01}$ (left panel), and the contributions of the shell
corrections calculated at the frozen densities and at incident
energies 95 MeV, 100 MeV and 105 MeV, respectively.}
\end{figure}
%%%%%%%%%%%%%%%%%%%%%%%%%%%%%%%%%%%%%%%%%%%%%%%%%%%%%%%%%%%%%%%%%%%%%%%%%

%%%%%%%%%%%%%%%%%%%%%%%%%%%%%%%%%%% figure 2 %%%%%%%%%%%%%%%%%%%%%%%%%%%%
\begin{figure}
\begin{center}
{\includegraphics*[width=0.8\textwidth]{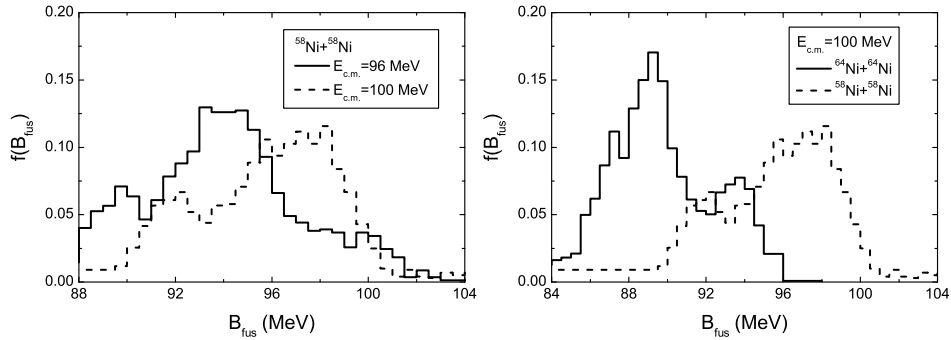}}
\end{center}
\caption{Distribution of the dynamical fusion barriers at incident
energies 96 MeV and 100 MeV in the center of mass frame (left panel)
and comparison of the systems $^{58}$Ni+$^{58}$Ni and
$^{64}$Ni+$^{64}$Ni (right panel).}
\end{figure}
%%%%%%%%%%%%%%%%%%%%%%%%%%%%%%%%%%%%%%%%%%%%%%%%%%%%%%%%%%%%%%%%%%%%%%%%%

%%%%%%%%%%%%%%%%%%%%%%%%%%%%%%%%%%% figure 3 %%%%%%%%%%%%%%%%%%%%%%%%%%%%
\begin{figure}
\begin{center}
{\includegraphics*[width=0.8\textwidth]{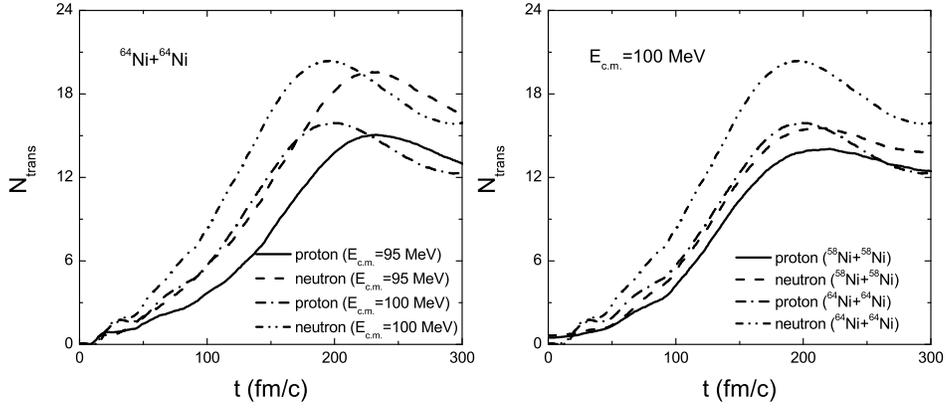}}
\end{center}
\caption{Nucleon transfer from projectile to target nucleus in the
neck region at different incident energies (left panel) and for
systems $^{58}$Ni+$^{58}$Ni and $^{64}$Ni+$^{64}$Ni (right panel).}
\end{figure}
%%%%%%%%%%%%%%%%%%%%%%%%%%%%%%%%%%%%%%%%%%%%%%%%%%%%%%%%%%%%%%%%%%%%%%%%%

%%%%%%%%%%%%%%%%%%%%%%%%%%%%%%%%%%% figure 4 %%%%%%%%%%%%%%%%%%%%%%%%%%%%
\begin{figure}
\begin{center}
{\includegraphics*[width=0.8\textwidth]{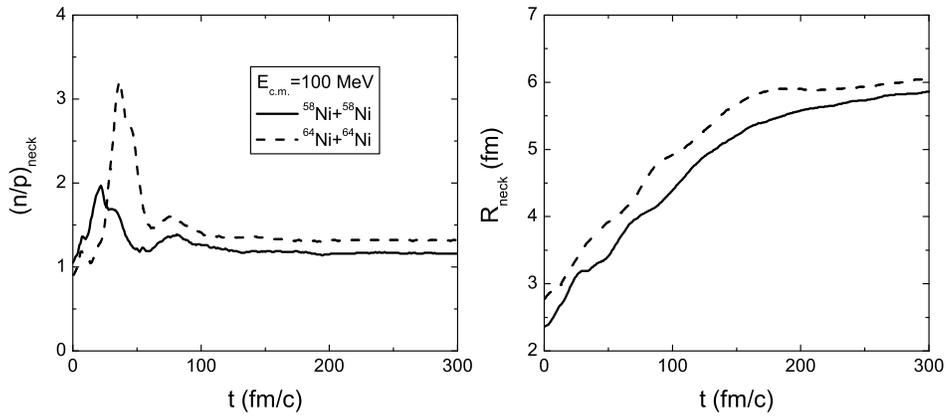}}
\end{center}
\caption{The ratio of neutron to proton in the neck region (left
panel) and the radius of the neck (right panel) as functions of the
evolution time at incident energy 100 MeV.}
\end{figure}
%%%%%%%%%%%%%%%%%%%%%%%%%%%%%%%%%%%%%%%%%%%%%%%%%%%%%%%%%%%%%%%%%%%%%%%%%

%%%%%%%%%%%%%%%%%%%%%%%%%%%%%%%%%%% figure 5 %%%%%%%%%%%%%%%%%%%%%%%%%%%%
\begin{figure}
\begin{center}
{\includegraphics*[width=0.8\textwidth]{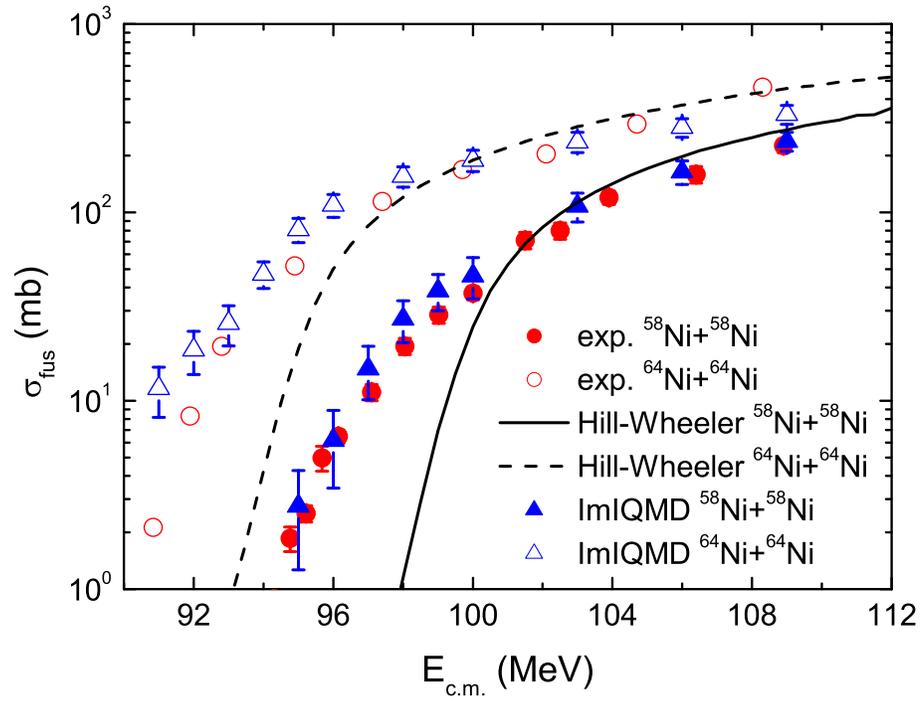}}
\end{center}
\caption{The calculated fusion excitation functions for the
reactions $^{58}$Ni+$^{58}$Ni and $^{64}$Ni+$^{64}$Ni, and compared
them with the Hill-Wheeler formula $\cite{Hi53}$ and the
experimental data $\cite{Be81,Ji04}$.}
\end{figure}
%%%%%%%%%%%%%%%%%%%%%%%%%%%%%%%%%%%%%%%%%%%%%%%%%%%%%%%%%%%%%%%%%%%%%%%%%

\end{document}